\documentstyle[aps,twocolumn,floats,prd,psfig]{revtex}

\newcommand{\mybib}[2]{\bibitem{#2}}
\newcommand{\ApJL}{Astrophys. J. Lett.}
\newcommand{\ApJ}{Astrophys. J.}
\newcommand{\PRL}{Phys. Rev. Lett.}
\newcommand{\PRD}{Phys. Rev. D}
\newcommand{\MNRAS}{MNRAS}

\newcommand{\aut}[2]{{#2.\ #1}}
\newcommand{\refs}[6]{#2, {\bf #3} {#4} (#5)}

\newcommand{\amp}{and }

\long\def\comment#1{}

\def\la{\hbox{ \raise.35ex\rlap{$<$}\lower.6ex\hbox{$\sim$}\ }}
\def\ga{\hbox{ \raise.35ex\rlap{$>$}\lower.6ex\hbox{$\sim$}\ }}

\def\W2{{\cal W}}

\newcommand{\wj}{\left(
                        \begin{array}{ccc}
                        l_1  &  l_2  & l_3 \\
                          0  &  0    &  0
                          \end{array}
                          \right)}
\newcommand{\wjm}{\left(
                         \begin{array}{ccc}
       l_1 & l_2  & l_3  \\
         m_1 & m_2  & m_3
                         \end{array}
                   \right)}

\newcommand{\bi}{B_{l_1 l_2 l_3}}

\newcommand{\deld}{\delta^{\rm D}}
\newcommand{\bn}{\hat{\bf n}}
\newcommand{\bm}{\hat{\bf m}}
\newcommand{\bl}{\hat{\bf l}}

\newcommand{\rad}{r}  
\newcommand{\da}{d_A} 

\newcommand{\Ylm}[1]{Y_{l_#1}^{m_#1}}
\newcommand{\Ylmn}{Y_{l}^{m}}
\newcommand{\alm}[1]{a_{l_#1 m_#1}}
\newcommand{\almn}{a_{l m}}

\begin{document}
\twocolumn[\hsize\textwidth\columnwidth\hsize\csname
@twocolumnfalse\endcsname

\title{Squared Temperature-Temperature Power Spectrum as a Probe of
the CMB Bispectrum}
\author{Asantha Cooray}
\address{
Department of Astronomy and Astrophysics, University of Chicago,
Chicago, IL 60637. \\ E-mail: asante@hyde.uchicago.edu}

\date{Submitted to PRD}

\maketitle

\begin{abstract}
It is now well known that mode-coupling effects associated with
certain secondary effects generate higher order correlations
 in cosmic
microwave background (CMB) temperature anisotropies, beyond the
two-point function.
In order to extract such a non-Gaussian signal at the three-point level,
we suggest a two-point statistic in the form of an
angular power spectrum  involving correlations between 
squared temperature and  temperature anisotropies.
This power spectrum contains compressed information from the
bispectrum and can be easily measured in data with the same
techniques that have been considered for the measurement of the
usual temperature-temperature anisotropy power spectrum.
We study the proposed power spectrum  resulting from
the non-Gaussian signal generated by correlations involved
with gravitational lensing angular deflections in CMB 
and the Sunyave-Zel'dovich (SZ)
effect due to large scale pressure fluctuations. 
Using the Planck frequency cleaned CMB and SZ maps, the
CMB$^2$-SZ power spectrum provides a direct estimate of the 
cross-power between lensing angular deflections and the SZ effect.
Through an optimal filter applied to the squared CMB map, 
the proposed statistic allows 
one to obtain all information from the lensing-SZ bispectrum. 
The observational measurement of the lensing-SZ cross-correlation 
is useful to understand the relation between large scale structure 
pressure and dark matter fluctuations.
\end{abstract}
\vskip 0.5truecm
]



\section{Introduction}

It is by now well accepted that the precision measurements of 
the cosmic microwave background (CMB) expected from upcoming 
experiments, especially MAP\footnote{http://map.nasa.gsfc.gov} and
Planck surveyor\footnote{http://astro.estec.esa.nl/Planck/; also, ESA
D/SCI(6)3.}, will provide adequate information for 
a precise measurement of cosmological parameters through the CMB 
temperature anisotropy power spectrum \cite{est}. 
In addition to a measurement of the angular
power spectrum, these experiments provide all-sky maps
across a wide range frequencies from $20$ GHz to $900$ GHz 
allowing the possibility to carry
out a large number of secondary sciences, which are
arguably as interesting as the primary goal. 
The possibilities for such studies with multifrequency CMB data 
include galactic foregrounds, secondary anisotropies such as the
Sunyaev-Zel'dovich (SZ; \cite{SunZel80}) effect resulting from
   the inverse-Compton scattering of CMB photons via hot gas in large
 scale structure \cite{Cooetal00} and far-infrared background (FIRB) 
due to high redshift dusty starforming galaxies \cite{Knoetal00}.    

The increase in sensitivity of these upcoming satellite missions 
and their wide-field coverage in many frequencies also raise the 
possibility that non-Gaussian signals in the CMB temperature 
fluctuations may be experimentally detected and studied in detail.
The deviations from Gaussianity in CMB
temperature fluctuations arise through two possibilities: the
existence of a primordial non-Gaussianity
due to initial conditions \cite{KomSpe00} 
and the creation of non-Gaussian 
fluctuations through either the imprint of 
non-linear growth of structures or mode-coupling effects 
by secondary sources of temperature
fluctuations \cite{CooHu00}. Note that in currently favored
adiabatic CDM models, the primary non-Gaussian contribution is
insignificant \cite{KomSpe00} and that main contributions in producing
a non-Gaussian signal comes from non-linear effects associated with
large
scale structure contributions to CMB and through various mode coupling
effects such as gravitational lensing.

As discussed in these papers, the detection of such non-Gaussian
signals at the three-point level 
is important for understanding inflation or large scale structure
clustering. The direct detection of such non-Gaussian signals, 
however, through a measurement of the 
full angular bispectrum, the Fourier analogue of the
three-point correlation  function, may be challenging.
Similar to measurements of the bispectrum in COBE \cite{Feretal98},
 it is likely that the future measurements will only be limited to
certain configurations of the bispectrum, such as
equilateral triangles in multipole space.

Keeping the possibility for experimental detections of
the non-Gaussian signals in mind, we considered alternative
statistics that include information from the higher order
level but can be extracted essentially through a modified two-point
correlation function. In the present paper, we discuss the
angular power spectrum associated with correlations between 
squared temperature and temperature.\footnote{
Additionally, the squared 
temperature-squared temperature power spectrum can be used
as a compressed measurement of the trispectrum. In Cooray, 
\cite{Coo01}, this is discussed as a possibility to extract the
non-linear kinetic SZ effect while in Hu \cite{Hu01}, this is
considered as a possibility to extract the lensing signal in CMB
data.} This power spectrum 
is essentially a compressed form of the bispectrum and can be 
computed using the same techniques that are well known with the
measurements of the usual temperature anisotropy power spectrum.
Thus, we do not expect	the measurement of the proposed
statistic to be affected by issues related to computation as in the
case of the full bispectrum. 

As an illustration of the astrophysical uses of the squared
temperature-temperature power spectrum, we consider the observational 
extraction of the bispectrum formed by non-linear mode coupling due to
gravitational lensing angular deflections in CMB data. Here, we
consider explicitly the correlation between lensing deflections and
the Sunyave-Zel'dovich effect \cite{SunZel80} due to inverse-Compton
scattering of CMB photons via hot electrons. Due to the spectral
dependence of the SZ effect when compared to CMB thermal fluctuations,
the SZ signal can be extracted from CMB fluctuations with
multifrequency data (see, \cite{Cooetal00}).
Here, we use such a frequency separation of CMB and SZ maps, in the
case of Planck mission, and
consider the measurement of the CMB$^2$-SZ power spectrum.
We show that this power spectrum is directly proportional to the
cross-correlation between SZ and lensing deflections. Thus,
with a measurement of the CMB$^2$-SZ power spectrum in Planck data, or
in any other multifrequency data set with reliable frequency
separation capabilities, one can directly probe the cross power
spectrum formed by dark matter, traced by lensing, and pressure,
traced by the SZ effect. The proposed method is one of the few ways to
obtain this information, which is important for proper understanding
of clustering of pressure relative to dark matter. 

Though we only discuss the CMB$^2$-SZ power spectrum as an application
of the squared temperature-temperature power spectrum there are
additional applications in both CMB and large scale structure studies.
Instead of measuring non-Gaussian statistics such as skewness or the
third moment, the proposed method can also be easily implemented for
weak lensing observations of the large scale structure using galaxy
shear data and for cross-correlation purposes between various probes
of large scale structure.

The layout of the paper is as follows.
In \S~\ref{sec:generalderiv}, we present a general calculation
 of the temperature$^2$-temperature angular power spectrum
and its relation to the angular bispectrum of temperature
anisotropies. In \S~\ref{sec:squared},
we illustrate the proposed angular power spectrum
through a calculation of the expected signal and noise
for the non-Gaussian signal produced through correlations between 
lensing angular deflection SZ effect.
We discuss our results in \S~\ref{sec:discussion} and conclude
with a summary.

\begin{figure}[t]
\centerline{\psfig{file=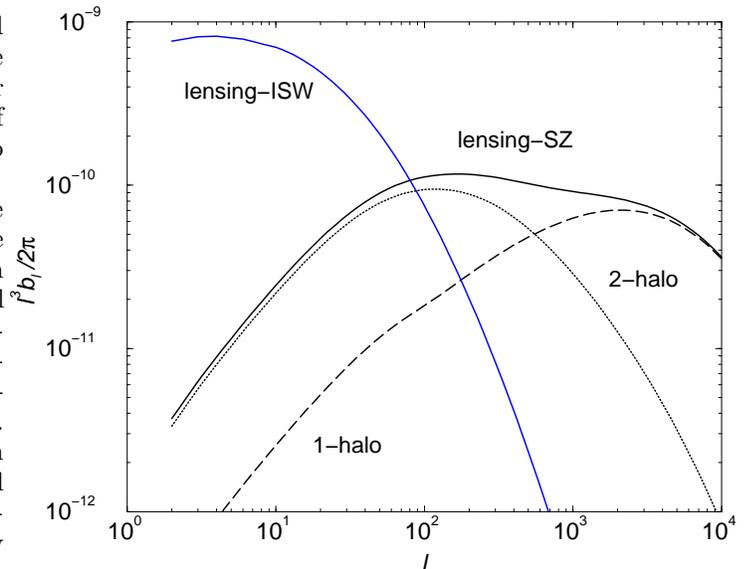,width=3.8in,angle=-90}}
\caption{Power spectrum for the correlation of 
lensing angular deflections and  the SZ effect, as calculated under
the halo description. As shown, most of the contributions to
the lensing-SZ correlation, as relevant for CMB, comes from
large angular scale correlations between halos, denoted by the 2-halo
term, and the mildly non-linear regime where contributions
from individual halos are important (1-halo term). 
For reference, we also show the correlation power spectrum
for lensing deflections and the integrated Sachs-Wolfe effect at
late times.}
\label{fig:bl}
\end{figure}

\section{Calculational Method}
\label{sec:generalderiv}

The bispectrum $\bi$
is the spherical harmonic transform of the three-point
correlation function just as the angular power spectrum $C_\ell$
is the transform of the two-point function.
In terms of the multipole moments of the
temperature fluctuation field $T(\hat{\bf n})$,
\begin{equation}
a_{lm} = \int d\bn T(\bn) \Ylmn {}^*(\bn)\,,
\label{eqn:spherical}
\end{equation}
the two point correlation function is given by
\begin{eqnarray}
C(\bn,\bm) &\equiv& \langle T(\bn) T(\bm) \rangle  \nonumber\\
           &=& \sum_{l_1 m_1 l_2 m_2} \langle \alm{1}^* \alm{2}
\rangle
               \Ylm{1} {}^*(\bn) \Ylm{2}(\bm)\,.
\label{eqn:twopoint}
\end{eqnarray}
Under the assumption that the temperature field is statistically
isotropic, the correlation is independent of $m$
\begin{eqnarray}
\langle \alm{1}^* \alm{2}\rangle = \deld_{l_1 l_2} \deld_{m_1 m_2}
        C_{l_1}\,,
\end{eqnarray}
and is called the angular power spectrum.
Likewise the three point correlation function is given by
\begin{eqnarray}
B(\bn,\bm,\bl) &\equiv& \langle T(\bn)T(\bm)T(\bl) \rangle \\
               &\equiv&
                \sum 
                \langle \alm{1} \alm{2} \alm{3} \rangle
                \Ylm{1}(\bn) \Ylm{2}(\bm)  \Ylm{3}(\bl)\,,\nonumber
\end{eqnarray}
where the sum is over $(l_1,m_1),(l_2,m_2),(l_3,m_3)$.
Statistical isotropy again allows us
to express the correlation in terms of an $m$-independent function,
\begin{eqnarray}
\langle \alm{1} \alm{2} \alm{3} \rangle  = \wjm \bi\,.
\label{eqn:3cumulants}
\end{eqnarray}
Here the quantity in parentheses is the Wigner-3$j$ symbol.
The orthonormality relation for Wigner-3$j$ symbol implies
\begin{eqnarray}
\bi = \sum_{m_1 m_2 m_3}  \wjm
                \langle \alm{1} \alm{2} \alm{3} \rangle \,.
\label{eqn:bispectrum}
\end{eqnarray}
 
The angular bispectrum, $\bi$, contains all the information available
in the three-point correlation function.  For example, the skewness,
the pseudocollapsed 
three-point function of \cite{Hinetal95}  and the
equilateral configuration statistic of \cite{Feretal98}  can all
be expressed as linear combinations of the bispectrum
terms (see \cite{Ganetal94} for explicit expressions and
\cite{Cooetal00} for an expression relating skewness in terms of
the bispectrum).

The quantity of interest here is the correlation between 
squared temperature and temperature, instead of the
usual temperature-temperature correlation. 
Following equation~\ref{eqn:twopoint},
we can write this  power spectrum as
\begin{eqnarray}
C^{2}(\bn,\bm) &\equiv& \langle T^2(\bn) T(\bm) \rangle  \nonumber\\
           &=& \sum_{l_1 m_1 l_2 m_2} \langle \alm{1}^{2} \alm{2}^*
\rangle
               \Ylm{1}(\bn) \Ylm{2} {}^*(\bm)\,.
\label{eqn:twopointsquared}
\end{eqnarray}
Here, note that $\alm{1}^2$ is the multipole moments of the squared
temperature field and not the square of the multipole
moment of the temperature field, $(\alm{1})^2$.
Through the expansion of the temperature
\begin{equation}
T(\bn) = \sum \almn \Ylmn(\bn),
\end{equation}
we can write
\begin{eqnarray}
a_{lm}^2 &=& \int d\bn T^2(\bn) \Ylmn {}^*(\bn) \nonumber \\
	 &=& 	\sum_{l_1 m_1 l_2 m_2} \alm{1} \alm{2} \int d\bn 
\Ylmn {}^*(\bn) \Ylm{1}(\bn) \Ylm{2}(\bn) \, .
\end{eqnarray}
We can now construct the power spectrum of 
squared temperature and temperature as 
\begin{eqnarray}
&&\langle a_{lm}^2 a_{l'm'}^* \rangle = C_l^{2} \deld_{l l'} \deld_{m m'} \nonumber \\
&=& \sum_{l_1 m_1 l_2 m_2} \langle \alm{1} \alm{2} a_{l'm'}^* 
\rangle  \int d\bn \Ylmn {}^*(\bn) \Ylm{1}(\bn) \Ylm{2}(\bn) \, .
\label{eqn:c2power}
\end{eqnarray}

Using the Gaunt integral
\begin{eqnarray}
\int d\bn
        \Ylm{1}
        \Ylm{2}
        \Ylm{3}
&=&
\sqrt{
        (2 l_1+1) (2 l_2+1) (2 l_3+1)\over 4\pi }\nonumber\\
&&\times
        \wj \wjm \,,  \nonumber \\
\label{eqn:harmonicsproduct}
\end{eqnarray}
and introducing the bispectrum from equation~\ref{eqn:3cumulants},
we can write the angular power spectrum of squared temperature and
temperature as
\begin{equation}
C_l^{2} =  \sum_{l_1 l_2} B_{l_1 l_2 l} w_{l_1,l_2} \left(
                         \begin{array}{ccc}
       l_1 & l_2  & l  \\
         0 & 0 & 0
                         \end{array}
                   \right)
\sqrt{
        (2 l_1+1) (2 l_2+1) \over 4\pi (2 l+1) } \, .
\label{eqn:finalform}
\end{equation}
In simplifying, we have made use of the fact that
\begin{equation}
\sum_{m_1 m_2} \left(
                         \begin{array}{ccc}
       l_1 & l_2  & l  \\
         m_1 & m_2  & m
                         \end{array}
                   \right)
\left(
                         \begin{array}{ccc}
       l_1 & l_2  & l'  \\
         m_1 & m_2  & m'
                         \end{array}
                   \right) = \frac{\deld_{l l'} \deld_{m m'}}{2 l+1}
\, .
\end{equation}

Equation~\ref{eqn:finalform}, is the expression of interest here.
This relates the angular bispectrum to the angular power
spectrum of squared temperature and temperature. 
In taking a general approach, we have introduced a filter, or window,
function in Fourier space of $w_{l_1,l_2}$. We will later discuss
detailed forms for the appropriate filter functions later.

Since the bispectrum is defined by a triangle in multipole space
with lengths of sides $(l,l_1,l_2)$, the $C_l^{2}$ power spectrum
essentially captures information, through a summation, 
associated with all triangular
configurations with one of the sides of length $l$. If apriori known
that certain triangular configurations contribute to the bispectrum
significantly, such as flattened triangles in the case of certain
secondary correlations (see, \cite{CooHu00}), one can
compute this sum by only restricting the multipoles of interest. This
is essentially what can be achieved with the introduction of
 an appropriate window, or a filter, function to the
squared temperature field.
Though the expression for the angular power spectrum involves a 
bispectrum, 
the experimental measurement is straightforward: one construct the
power spectrum by squaring the temperature field, in real space,
 and using the Fourier transforms of  squared temperature values and
the  temperature field, with any filtering functions, when
necessary.

We will now discuss the
measurement of the power spectrum using Planck data for the
non-Gaussian signal involved through the correlation between lensing
angular deflections and the SZ effect. This bispectrum
has the high cumulative signal-to-noise out of all other possible
bispectra involving secondary effects and CMB temperature
\cite{CooHu00}.  For the
illustration of our results we use the currently favored $\Lambda$CDM
cosmological model. The parameters for this model
are $\Omega_c=0.30$, $\Omega_b=0.05$, $\Omega_\Lambda=0.65$, $
h=0.65$, $Y_p = 0.24$, $n=1$, COBE normalization 
$\delta_H=4.2 \times 10^{-5}$ \cite{BunWhi97}.
This model has mass fluctuations on the $8 h$ Mpc$^{-1}$
scale in accord with the abundance of galaxy clusters
$\sigma_8=0.86$ \cite{ViaLid99}.  

\section{CMB$^2$-SZ power spectrum}
\label{sec:squared}

\begin{figure}[t]
\centerline{\psfig{file=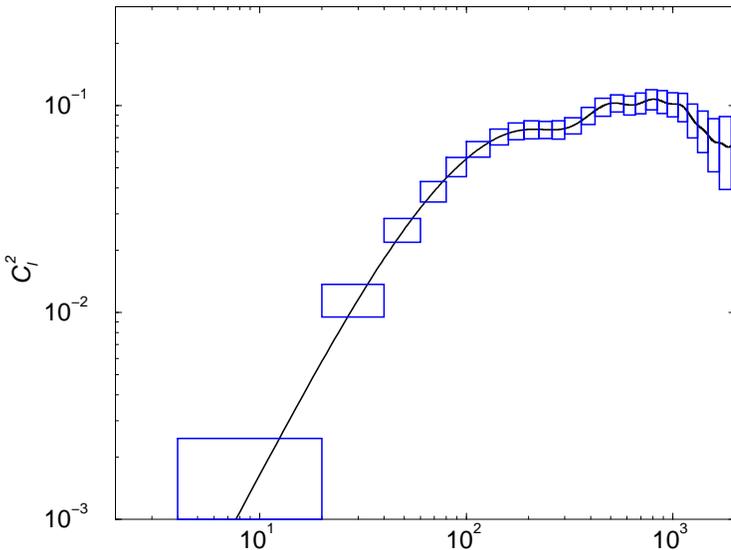,width=3.8in,angle=-90}}
\caption{The angular power spectrum of
CMB$^2$-SZ due to the correlation of SZ effect and gravitational
lensing in CMB. The band power errors are for the Planck mission using the noise calculation for multifrequency separation of SZ and CMB 
effect from Planck data.}
\label{fig:cl}
\end{figure}

\begin{figure*}[t]
\centerline{\psfig{file=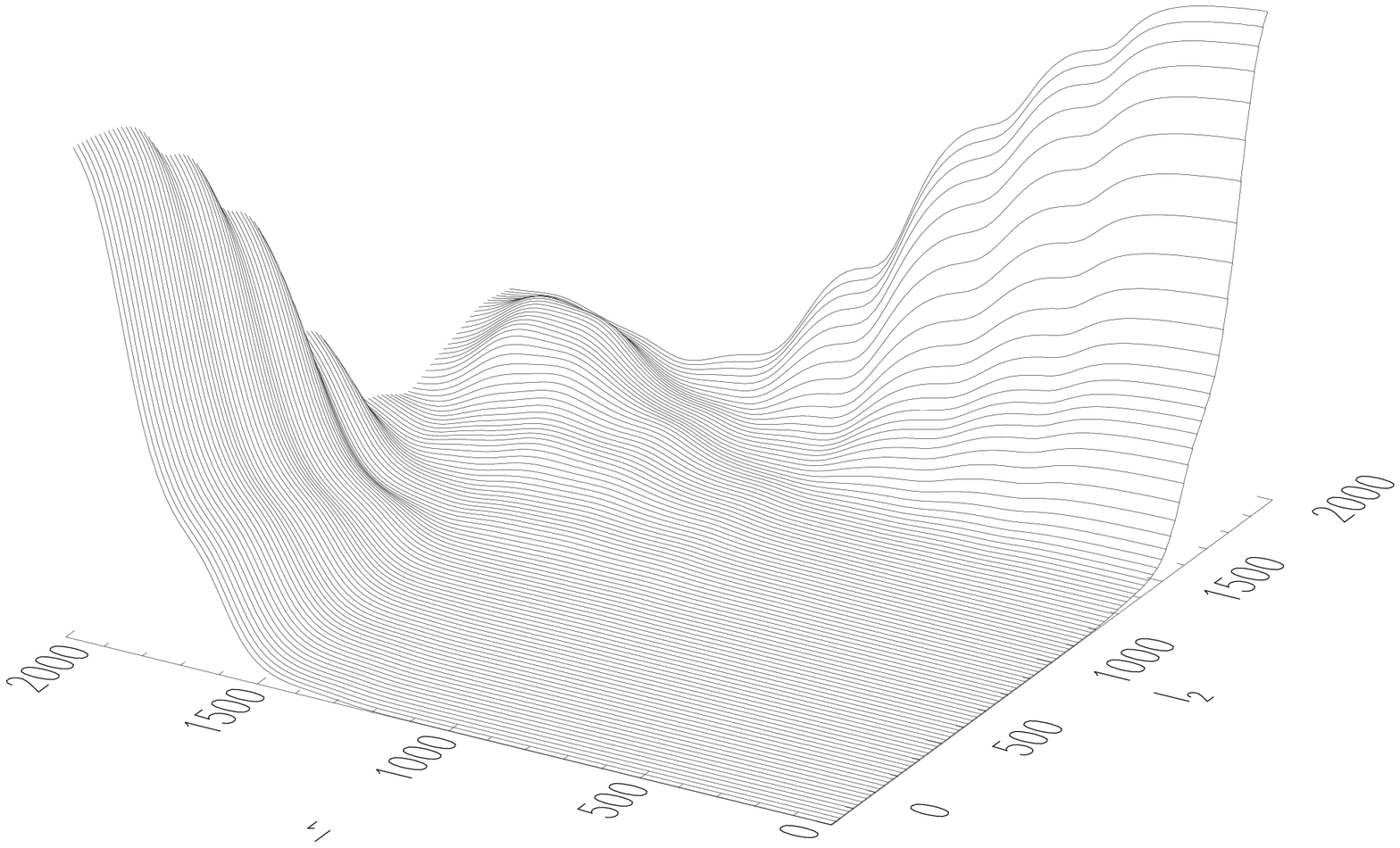,width=3.8in,angle=90}
\psfig{file=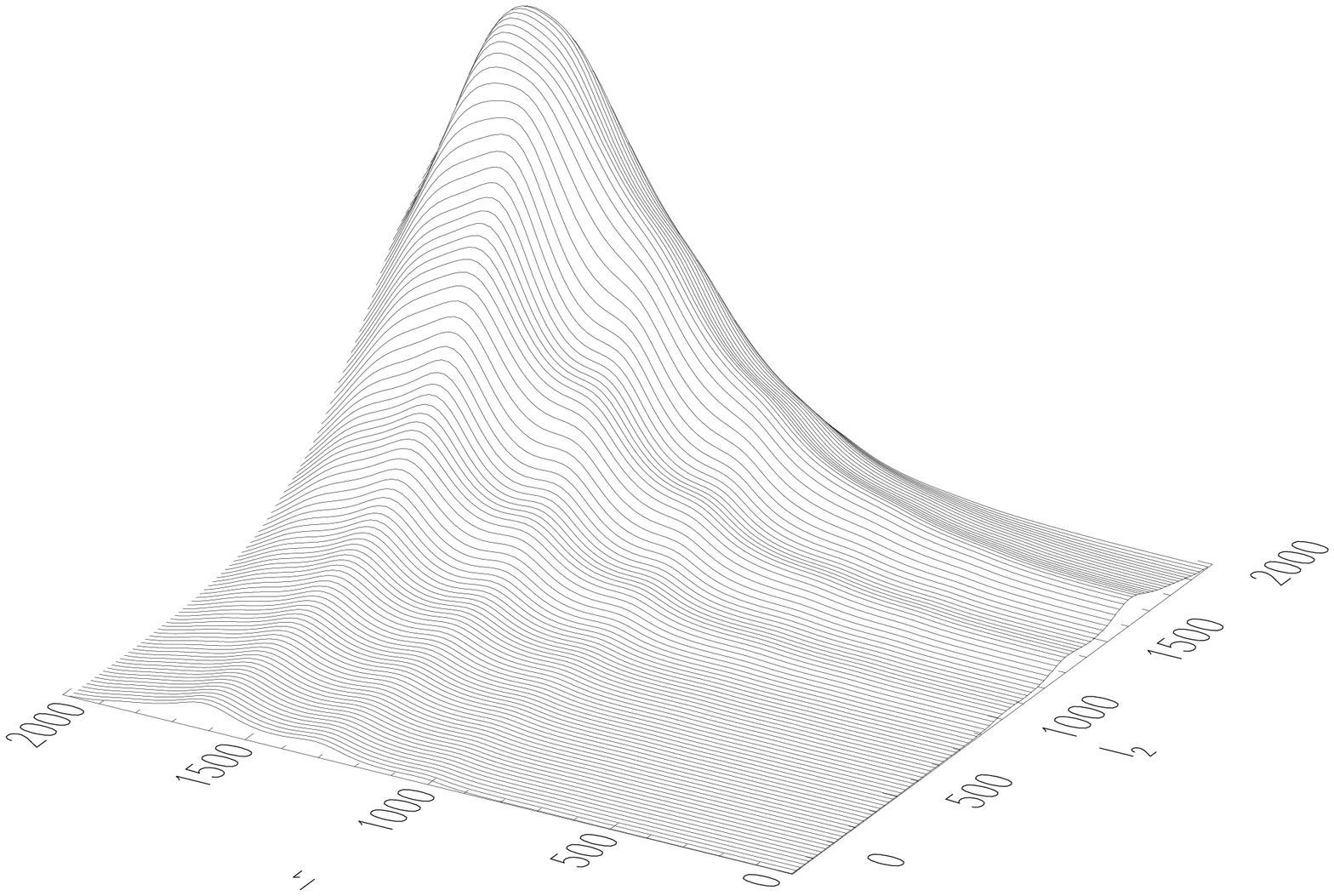,width=3.8in,angle=90}}
\caption{Optimal filter function for the CMB$^2$-SZ power spectrum,
as a function of $l_1$ and $l_2$. In left, the filter is when $l=100$
and
right is when $l=2000$ corresponding to large and small angular scales
respectively. The optimal filter behaves 
such that it removes excess noise at multipoles less than $\sim$ 1500
and extracts sensitivity of lensing in the damping tail of the power spectrum.}
\label{fig:filter}
\end{figure*}

We refer the reader to \cite{CooHu00} for a full derivation
of the bispectra due to correlations between lensing angular
deflections in CMB and secondary effects due to large scale 
structure. Following these derivations, we can write the
bispectrum as
\begin{eqnarray}
&& B_{l_1,l_2,l} = - \left(
                         \begin{array}{ccc}
       l_1 & l_2  & l  \\
         0 & 0 & 0
                         \end{array}
                   \right)
\sqrt{ \frac{(2l_1 +1)(2 l_2+1)(2 l+1)}{4 \pi}}
\nonumber \\
&\times&
\left[f_{l_2,l_1,l} C_{l_1}^{\rm CMB}b_{l}+ {\rm
Perm.}\right]\,,
 \nonumber \\
\label{eqn:cmbbi}
\end{eqnarray}
where
\begin{equation}
f_{l_a,l_b,l_c} =
\left[\frac{l_a(l_a+1)-l_b(l_b+1)-l_c(l_c+1)}{2}\right] \, ,
\end{equation}
and $b_l$ is the cross-correlation power spectrum involved with
lensing angular deflections and a secondary source of anisotropies
that trace the large scale structure and effectively correlates with
lensing potentials (e.g., integrated Sachs-Wolfe effect
\cite{SacWol67}, linear Doppler effect, SZ effect).
Here, we consider the case of SZ since, as discussed in
\cite{CooHu00}, the correlation between lensing and other two effects
does not lead to a bispectrum with significant  signal-to-noise.

If the two maps involved with the construction of the
squared temperature-temperature power spectrum is the same (e.g., CMB alone),
then the permutation in equation~\ref{eqn:cmbbi}
leads to a total of 6 terms in the ordering of
$l_1$, $l_2$ and $l$. 
When the bispectrum is measured with  two independent maps, e.g., CMB and SZ
through frequency cleaning,  the permutation in equation~\ref{eqn:cmbbi} 
involves an additional term with the
replacement of $l_1$ by $l_2$. 

We now primarily consider the case of two
independent maps with Planck, and thus, we can write the CMB$^2$-SZ
power spectrum as
\begin{eqnarray}
C_l^{2} &=& b_l \sum_{l_1 l_2}   \left(
                         \begin{array}{ccc}
       l_1 & l_2  & l  \\
         0 & 0 & 0
                         \end{array}
                   \right)^2
        {(2 l_1+1) (2 l_2+1) \over 4\pi } w_{l_1,l_2}^2 \nonumber \\
&\times& \left[C_{l_1}^{\rm CMB}f_{l_2,l_1,l}
+C_{l_2}^{\rm CMB}f_{l_1,l_2,l}\right] \, .
\label{eqn:cl2finalform}
\end{eqnarray}
Here,
$C_l^{\rm CMB}$ is the unlensed primary CMB contribution. 

The proposed use of the CMB$^2$-SZ power spectrum is essentially
following. One can directly measure $C_l^2$ in the CMB and SZ data 
and since information related to $C_l^{\rm CMB}$ also directly comes
from data, one can construct the quantity $b_l$ which is directly
proportional to $C_l^2$. As written in equation~\ref{eqn:cl2finalform}, 
in the case relevant to the CMB$^2$-SZ power spectrum,
the cross-correlation between lensing deflections and the SZ 
effect 
\begin{equation}
b_l =  \int d\rad \frac{
        W^{\rm SZ}(r) W^{\rm len}(k,r)|_{k=l/\da}}{\da^2}
        P_{\Pi \delta}({ l \over \da};r) \,.
\label{eqn:bl}
\end{equation}
Here, we have utilized the Limber approximation \cite{Lim54} by
setting $k=l/\da$, $P_{\Pi \delta}$ is the pressure-dark matter
cross power spectrum, $r$ is the comoving radial distance
and $\da$ is the comoving angular diameter distance.
The window functions for the SZ  and lensing deflections are
\begin{equation}
W^{\rm SZ}(r) = g(x) \frac{k_B \sigma_T \bar{n}_e}{a(r)^2 m_e c^2}
\label{eqn:wsz}
\end{equation}
and 
\begin{equation}
W^{\rm len}(k,r) = -3 \frac{\Omega_m}{a(r)} 
\left(\frac{H_0}{k}\right)^2 \frac{\da(r_s-r)}{\da(r) \da(r_s)} \, ,
\label{eqn:wlens}
\end{equation}
respectively. 
In Eq.~\ref{eqn:wsz}, $\bar{n}_e$ is the mean density of electrons
today, $\sigma_T$ is the Thomson cross-section  and $g(x) =
x \coth (x/2)-4$ with $x=h\nu/k_B T_{\rm CMB}$ is the spectral shape
of the SZ effect. At Rayleigh-Jeans part of the CMB, $g(x)=-2$.
As discussed in \cite{Cooetal00}, an experiment such as Planck with
observations over a wide range of frequencies  can
separate out SZ contributions based on the spectral signature, $g(x)$
relative to primary CMB and other contributors.
For this paper, we will consider Planck in estimating signal-to-noise
and will use the noise power spectra calculated in \cite{Cooetal00}
for the Planck SZ and CMB maps.

In order to assess how well Planck CMB and SZ maps can be used for the
purpose of constructing $b_l$, we need to consider an estimate for it,
as well as, estimates for $C_l^{\rm CMB}$ and $C_l^{\rm SZ}$;
The latter is required when estimating the signal-to-noise for
possible detections. We use CMBFAST of \cite{SelZal96} and compue
$b_l$ and $C_l^{\rm CMB}$ analytically using a semi-analytical
description of pressure fluctuations in the large scale structure
involving the halo approach
\cite{Sel00}. To calculate $b_l$ we need the
pressure-dark matter cross-power
spectrum while $C_l^{\rm SZ}$ depends on the pressure-pressure power
spectrum. The power spectra are introduced and discussed in detail in
\cite{Coo00,Coo01}  and we refer the reader to these two papers for
for further details on this approach. The basic assumption of the halo
model is that large scale structure dark matter distribution
 can be viewed as a collection of dark matter halos with a mass
function following Press-Schechter \cite{PreSch74} mass function or
variants and a dark matter distribution in each halo following NFW profile of
\cite{Navetal96} or other descriptions. The halos are biased relative
to the linear density field following the description of
\cite{Moetal97}. With clustering of dark matter described through such
an approach, any other physical property of the large scale structure
can be easily described through the relation between that property and
dark matter. For example, in the case of pressure, we can gas
distribution in each halos follow hydrostatic equillibrium with dark
matter.  As discussed in \cite{Coo01}, 
such an approach through the halo model provides a reliable semi-analytic
approach to model non-linear large scale structure clustering and its
predictions, in the case of pressure as relevant for the SZ effect and
lensing-SZ correlations, are consistent with numerical simulations (e.g.,
\cite{RefTey01}).  

In figure~\ref{fig:bl}, we show the correlation
between lensing deflections and SZ effect. The contributions are
broken to singlem 1-halo, and 2-halo term involving correlations between and
within halos. As shown, for multipoles of interest, most of the
correlation comes from the 2-halo term involving large scale
correlations between halos, though at multipoles $\sim$ few hundred, the
contribution from single halo term cannot be ignored. This is
the mildly non-linear regime where contributions from large scale
correlations between halos and the halo profiles become important.

Following the correlation between lensing and SZ shown in
figure~\ref{fig:bl}, in figure~\ref{fig:cl}, we show the
$C_l^2$ power spectrum due to the relevant bispectrum.
We now discuss the calculation of errors shown in figure~\ref{fig:cl}.

\subsection{Signal to Noise}

To establish the signal-to-noise for the detection of the lensing-SZ
bispectrum through the squared temperature-temperature power spectrum, we calculate the covariance associated with the proposed statistic
following \cite{Coo01}. Here, we assume that the squared temperature
and temperature power spectrum will be measured with two maps
involving 
CMB and SZ with squared temperature measurement from the CMB map 
and a single temperature measurement from SZ.
We can write the error on each of the power spectrum measurements
associated  with the proposed statistic as
\begin{equation}
\Delta C_l^{2} = \sqrt{\frac{1}{f_{\rm sky} (2l+1)}}
\left[\left(C_l^{2}\right)^2 +C_{l,tot}^{2 \rm CMB}
C_{l,tot}^{\rm SZ}\right]^{1/2} \, ,
\label{eqn:error}
\end{equation}
while the total signal-to-noise is
\begin{equation}
\frac{S}{N} = \left[ f_{\rm sky } \sum_l (2l+1)
\frac{\left(C_l^2\right)^2}{\left(C_{l}^2\right)^2 +C_{l,tot}^{2 \rm CMB}
C_{l,tot}^{\rm SZ}}\right]^{1/2} \, ,
\label{eqn:sn}
\end{equation}
with the fraction of sky covered by $f_{\rm sky}$.

Here, the propose squared temperature and temperature
angular power spectrum  is $C_l^2$ 
and $C_{l,tot}^{2 \rm CMB}$ is the squared 
temperature power spectrum in CMB data alone,
\begin{eqnarray}
&&C_{l,tot}^{2 \rm CMB} =  \nonumber \\
&&\sum_{l_1 l_2} 2 C_{l_1,tot}^{\rm CMB} C_{l_2,tot}^{\rm CMB} w_{l_1,l_2}^2
\wj^2 \frac{ (2l_1+1) (2l_2+1)}{4 \pi} \, .
\end{eqnarray}

In calculating the noise, following \cite{Kno95},
we also include relevant detector
and frequency separation noise --- for the SZ map --- by
introducing $C_{l,tot}^{\rm CMB} =C_{l}^{\rm CMB} + C_{l}^{\rm S}+ 
C_{l}^{\rm N}$, with $S$ denoting other secondary contributors
in the CMB map (e.g., lensing) and $N$ is the noise contribution.
There is an additional term here involving the CMB trispectrum which
leads to the covariance between multipoles. We
ignore it here given that there is no significant non-Gaussian
component to the primary anisotropies especially in the adiabatic CDM
models \cite{KomSpe00}, which are more consistent with current
observations of large scale structure and CMB. 
The next important contribution to the covariance of CMB comes from
secondary effects with a thermal spectrum. As discussed in Hu
\cite{Hu01}, gravitational lensing of CMB, 
which is the most important secondary effect that  leads to non-Gaussian
correlations, does not produce a significant covariance out to $l$ of
2000 as relevant for the Planck mission. Thus, we ignore the
presence of a non-Gaussian trispectrum in our variance estimates and
only consider Gaussian contributions.
At further small angular scales, in addition to lensing,
other secondary effects with a CMB thermal 
spectral signature, such as kinetic SZ effect, 
can produce a trispectrum in CMB temperature  data (see,
\cite{Coo01}).

Note that since we consider frequency separated maps, there is 
no trispectrum contribution 
to the covariance of the proposed statistic from the SZ
effect and contribution to the variance comes only from the
SZ power spectrum $C_l^{\rm SZ}$. We calculated this power spectrum
again using the halo model following \cite{Coo01}.
If the measurement is to be done in a map which is 
not frequency cleaned with components separated out, 
then the trispectrum involved with SZ will also contribute
to the covariance.

\subsection{Filter functions}
\label{sec:filter}

Here, we consider two possibilities for the filtering of
unnecessary noise in the $T^2$ map. A straightforward filter is to
essentially remove the excess low multipole noise and we achieve
this by choosing
\[ \mbox{$w_{l_1,l_2}$} = \left\{ \begin{array}{ll}
        1 & \mbox{$l_1,l_2 > l_{\rm cut}$} \\
        0 & \mbox{$l_1,l_2 < l_{\rm cut}$}
        \end{array}
              \right. \]
where we set the $l_{\rm cut}$ to be 1000. Note that in the case of Planck,
instrumental noise degrades the information beyond a $l$ of 2000.

\begin{figure}[!h]
\centerline{\psfig{file=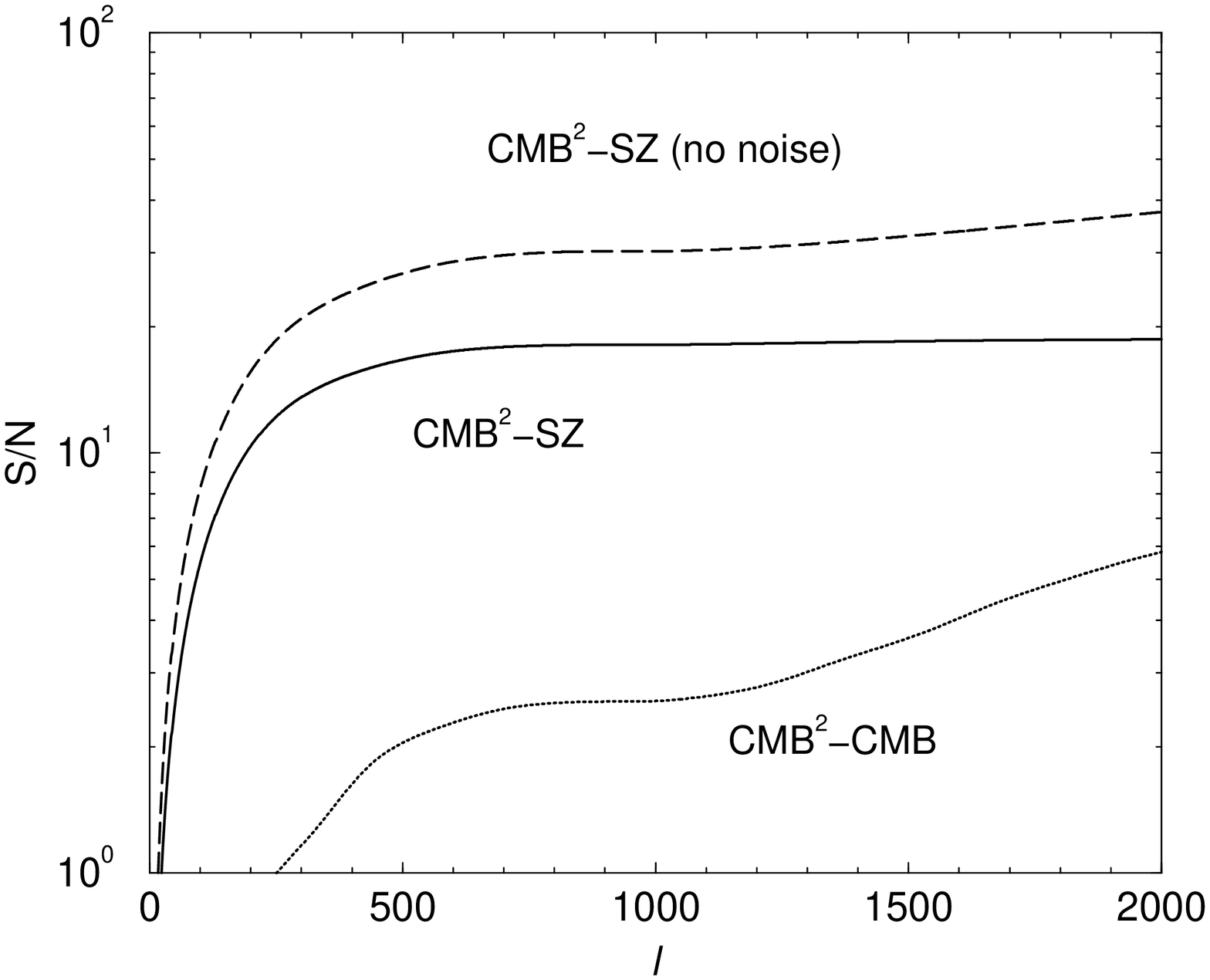,height=3in,width=3.7in,angle=-90}}
\centerline{\psfig{file=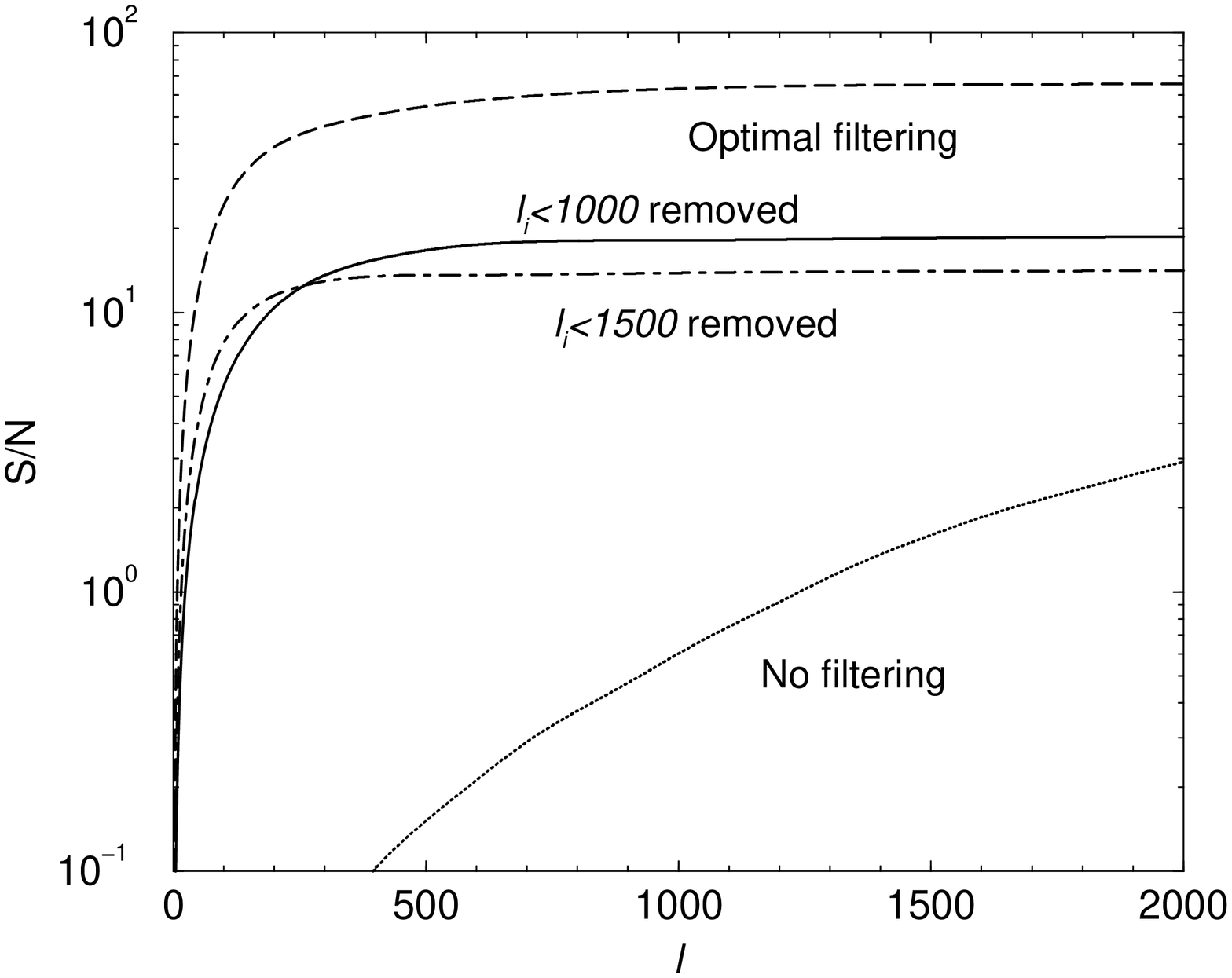,height=3in,width=3.7in,angle=-90}}
\caption{The cumulative signal-to-noise for the detection of
temperature$^2$-temperature power spectra in Planck data. The top
figure is using the filtering scheme where we remove excess noise at
multipoles $l_1$ and $l_2$ of less than 1000. The 
shown curves are for with no SZ separation ({\it dotted}),
with SZ separation ({\it solid}), and perfect SZ separation with
no noise contribution to both SZ and CMB maps ({\it dashed}).
In the realistic case of SZ separation, the total cumulative
signal-to-noise is of order $\sim$ 20. The bottom panel compares
the signal-to-noise for Planck with no filtering (dotted line) and
optimal filtering (long dashed line). Optimal filtering leads to a
cumulative signal-to-noise of $\sim$ 70.}
\label{fig:sn}
\end{figure}

Additionally, we consider an optimal filtering scheme which
maximizes the $(S/N)^2$ in equation~\ref{eqn:sn}.
Analytically, this is equivalent to taking the derivative of
the expression in equation~\ref{eqn:sn} with respect to $w_{l_1,l_2}$
and solving  for the value which maximizes the signal-to-noise.
We found this to be when
\begin{equation}
w_{l_1,l_2} = \frac{\left[C_{l_1}^{\rm CMB}f_{l_2,l_1,l}
+C_{l_2}^{\rm CMB}f_{l_1,l_2,l}\right]}{2 C_{l,tot}^{\rm CMB}
C_{l_2,tot}^{\rm CMB}} \, ,
\end{equation}
with an appropriate normalization such that $\sum_{l_1 l_2} w_{l_1,l_2}=1$.
This optimal filter  is equivalent to the one introduced by
Hu \cite{Hu01} for the estimator of the deflection angles in CMB using
a quadratic statistic. Note that the optimal filter is independent of the
correlation power between lensing deflections and the SZ effect, however,
depends on the details of mode coupling associated with lensing in CMB
through the $f$ terms. One can easily construct the filter through
apriori knowledge on the CMB power spectrum and noise properties.

When the  optimal filter is applied to the squared temperature map,
$C_{l,tot}^{2 \rm CMB} = C_l^2$, and this leads to a simplified
expression for  the signal-to-noise of the CMB$^2$-SZ  power spectrum
\begin{eqnarray}
\left(\frac{S}{N}\right)^2 &\approx& f_{\rm sky } \sum_{l l_1 l_2} 
   \left(
                         \begin{array}{ccc}
       l_1 & l_2  & l  \\
         0 & 0 & 0
                         \end{array}
                   \right)^2
        {(2 l_1+1) (2 l_2+1) \over 4\pi } \nonumber \\
&\times& \frac{\left[C_{l_1}^{\rm CMB}f_{l_2,l_1,l}
+C_{l_2}^{\rm CMB}f_{l_1,l_2,l}\right]^2 b_l^2}{2 C_{l_1,tot}^{\rm
CMB} C_{l_2,tot}^{\rm CMB} C_{l,tot}^{\rm SZ}} \nonumber \\
&=& f_{\rm sky} \sum_{l l_1 l_2} \frac{B^2_{l_1 l_2 l}}{2 C_{l_1,tot}^{\rm
CMB} C_{l_2,tot}^{\rm CMB} C_{l,tot}^{\rm SZ}} \, .
\label{eqn:bisn}
\end{eqnarray}

As written above, the total signal-to-noise 
is now equal to exactly the total amount present in the full bispectrum
formed by correlations between gravitational lensing deflections in
CMB and the SZ effect (see, \cite{CooHu00}). 
Thus, the optimal filter allows one to capture, at the two point level
through basically a power spectrum,
all information contained within the bispectrum at the three point
level; this happens with no loss in signal-to-noise. 
The temperature-gradient divergence statistic
introduced by \cite{Hu01} allows this to be carried out directly; 
with appropriate optimal filtering suggested above, this is
equivalent to the CMB$^2$-SZ approached outlined here.

In figure.~\ref{fig:filter}, we show the surface of the optimal filter
as a function of $l_1$ and $l_2$ when $l_3=100$ and 2000,
respectively.
As shown, the optimal filter behaves such that one only used
information in multipole range  with $l_1$ and $l_2$ $\gtrsim 1500$. 
This behavior is the reason why that a simple filter with the behavior
of a step-function in $l_1$ and $l_2$ essentially allows us to capture
some of the information present in the bispectrum. The detailed
behavior of the optimal filter function, especially information related to
lensing  present in the peaks and valleys of the CMB power spectrum,
however, limits the information that can be captured by a simple
filter. The optimal filter behaves such that it effectively extracts
all information related to lensing from the damping tail of the CMB
power spectrum.

\section{Discussion}
\label{sec:discussion}

In figure~\ref{fig:bl}, we show the correlation between
SZ effect and lensing angular deflections. The cross angular
power spectrum is weighted by a factor of $l^3$ to 
highlight the multipole range which is important for the
CMB bispectrum due to gravitational lensing-secondary correlations.
For comparison, in the same plot, we show the correlation power
between lensing deflections and the integrated Sachs-Wolfe effect
(ISW; \cite{SacWol67}) at late times. We refer the reader to
Cooray \& Hu \cite{CooHu00} for a detailed discussion of the bispectra 
produced through correlations between gravitational lensing angular
deflections and secondary effects. In the case of the lensing-SZ,
note that we have updated the calculations presented therein
using a detailed description of large scale structure pressure
fluctuations instead of the assumption that pressure traces dark
matter (see, \cite{Coo00,Coo01} for a discussion on the 
computation of
statistics related to large scale structure pressure);
The assumption of pressure traces dark matter, in general, leads to an
slight overestimate of the correlation between lensing and SZ.

The lensing-SZ correlation behaves such contributions
relavant for the CMB bispectrum, at multipoles less than 100,
 comes from interhalo large scale 
correlations, which we defined as the 2-halo term in 
\cite{Coo00,Coo01}, involving the linear density field.
The mildly non-linear regime, at multipoles around few hundred, is
described through the combination of large scale correlations and the
interhalo correlations through the profiles denoted by the 1-halo term.
This behavior is consistent with the fact that angular deflections in CMB 
is a large scale phenomena with an angular coherence of the deflection
angle of $\sim$ 10 degrees (see, \cite{HuCoo00}).
Though lensing windown function peaks at a redshift of $\sim$ 3,
at half the angular diameter distance to last scattering, the
non-linear contributions to the lensing-SZ correlations comes
primarily from the SZ effect which has a window function that
effectively peaks at very low redshifts; overall, contributions to the
lensing-SZ correlations comes over a wide-range of redshifts and the
low redshift behavior of the SZ effect makes the contribution
from the 1-halo term important at small angular scales.

Note that most of the contributions come at multipoles greater than
few hundred, with a tail  continues to multipoles of thousands.
Since CMB itself dominates low multipoles, its contribution
to the variance is more important at multipoles less than $\sim$
1000. This is the same reason why the lensing-ISW effect, even with
a significantly higher correlation at very large angular scales,
leads to considerably smaller signal-to-noise for the bispectrum
and thus, to the squared temperature-temperature power spectrum.

\begin{figure}[t]
\centerline{\psfig{file=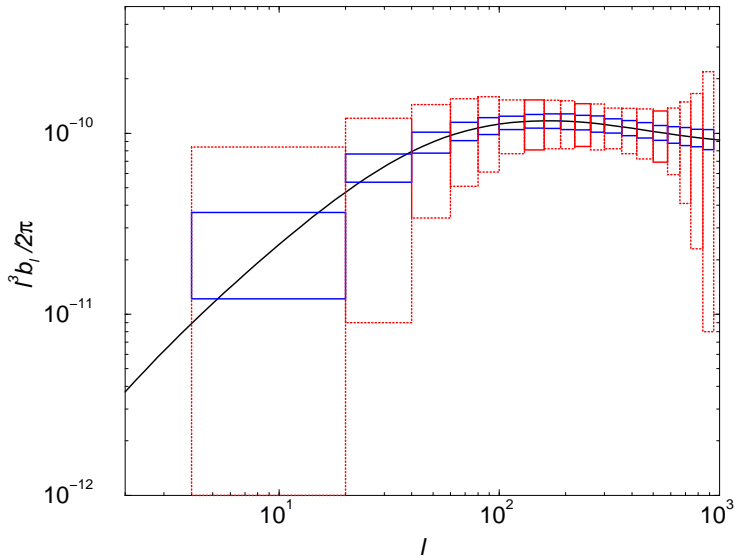,width=3.8in,angle=-90}}
\caption{Power spectrum for the correlation of 
lensing angular deflections and  the SZ effect, as calculated under
the halo description. The band power errors are from the measurement
of CMB$^2$-SZ power spectrum in Planck with frequency cleaned CMB and
SZ maps with a filtering scheme where noise is removed at multipoles
less than 1000 (dotted) and using the optimal filtering (solid). 
The experimental measurement of the SZ-lensing deflections
cross power spectrum is important for the understanding of relation
between  large scale pressure and dark matter fluctuations.}
\label{fig:blerror}
\end{figure}

In figure~\ref{fig:cl}, we show the angular power spectrum of
CMB$^2$-SZ power spectrum using the halo model description of $b_l$.
We computed the power spectrum band errors by using the 
optimal filtering scheme discussed  in \S~\ref{sec:filter} and
shown in figure~\ref{fig:filter} and using
equation~\ref{eqn:error} to compute errors. 
Here, we consider the Planck measurement of
the SZ effect and use the variance power spectrum calculated in
\cite{Cooetal00}. We also include the relevant beam/detector 
noise in the CMB map. For secondary contributors to the noise, we
include contribution from lensing and non-linear kinetic SZ following
\cite{HuCoo00} and \cite{Coo01}. The effect of the optimal filter is
to
remove the noise associated with low multipoles primarily from the
intrinsic CMB  anisotropies, while extracting information related to
lensing associated with peaks and valleys in the CMB power spectrum as
well as small scall information coming from the modification of a CMB
gradient on the sky.  As shown, Planck derived CMB and SZ maps can
 be clearly used to detect the CMB$^2$-SZ non-Gaussian bispectrum  due to
lensing at the  two-point level. 

In figure~\ref{fig:sn}, we show the cumulative signal-to-noise
associated with this detection. Here, we consider both the optimal
filtering as well as our crude filter where we removed all information
at multipoles less than 1000. In the top panel, we consider this
latter scenario. Using the frequency
cleaned maps, the signal-to-noise ranges up to $\sim$ 20, 
while with no frequency
cleaning we only find a value around 6. If Planck were to
allow perfect separation of SZ and CMB and with no noise 
contributions to both SZ and CMB maps, and with only cosmic 
variance, one finds a total signal-to-noise of $\sim$ 60. 
This increase in signal-to-noise primarily comes from multipoles
greater than 1000 and is due to the decrease in the noise associated
with the SZ map. Clearly, improving the separation of SZ from CMB and
decreasing the associated noise can lead to a significantly better
detection of the lensing-SZ correlation through the
CMB$^2$-SZ power spectrum. 

In the bottom panel of figure~\ref{fig:sn}, we compare cumulative
signal-to-noise with no filtering, optimal filtering and with
our simple noise removal at $l \lesssim 1000$ and $l \lesssim 1500$. 
Here again, we consider
the Planck CMB and SZ maps. With no filtering,
cumulative signal-to-noise is of order $\sim$ 3, while with optimal
filtering, one obtains a total signal-to-noise of $\sim$ 70. With the
simple noise removal at $l$ of 1000, 
one has only gained a factor of $\sim$ 6, while
this is still a factor of $\sim$ 3.5 less than optimal case. 
As shown in figure~\ref{fig:filter}, when large scales are considered,
the filter removes excess noise out to $l_1,l_2$ of 1500. Thus,
we introduced simple noise removal out to $l$ of 1500. As shown
in figure~\ref{fig:sn} (bottom panel) with a dot-dashed line,
even though there is an increase in signal-to-noise at low $l$ values,
consistent with the behavior of the optimal filter function, one
finds less cumulative signal-to-noise at small angular scales.
Such differences are primarily due to the detailed behavior of the optimal
filter. With a simple filter, no information is obtained from
multipoles that is associated with peaks and valleys of the CMB power
spectrum, while related information is contained within the optimal
filter. Though one can obtain relatively modest measurement of the
CMB$^2$-SZ power spectrum through crude filtering, an optimal
filter, such as the one suggested here, is highly recommended to
exploit the full potential of the measurable signal.

In addition to the CMB$^2$-SZ power spectrum, or a constructing
through the temperature-gradient divergence statistics of \cite{Hu01}, an 
alternative technique to construct the lensing-SZ correlation involves
the gradient  statistic of \cite{ZalSel99}. This technique utilizes 
  gradient maps and an analysis similar to polarization to construct a
map that is directly proportional to weak lensing convergence. With such a 
statistic, and following the procedure in \cite{ZalSel99,PeiSpe00},
we find a cumulative signal-to-noise of $\sim$ 35, which is better
than the simple filter but roughly a factor of 2 worse than the
optimal filter considered here.

Given that the proposed statistic involving CMB$^2$-SZ power spectrum
is directly proportional to the cross-power spectrum between lensing
angular deflections and the SZ effect, our error estimates on $C_l^2$
can be converted to those of $b_l$. In figure~\ref{fig:blerror}, we
show the band power errors on $b_l$ using the Planck estimates for
errors on CMB$^2$-SZ power spectrum using the simple filter (dashed
line)
and the optimal filter (solid lines).  As shown, Planck allows a clear
detection of the cross-correlation between SZ and lensing
deflections; the detection is very clear with the
optimal filter. The experimental measurement of this
 correlation power spectrum is preferred since it contains important 
cosmological and astrophysical information related to large scale
fluctuations of pressure and dark matter.
When combined with SZ-SZ power spectrum and the
lensing-lensing power spectrum either from large scale structure weak
lensing or CMB directly, this lensing-SZ power spectrum will  provide us
knowledge on the correlation between pressure and dark matter.
The proposed squared temperature-temperature statistic involving
Planck CMB and SZ maps, with appropriate filtering,
 is clearly one of the few ways to obtain this 
information.

\acknowledgments
We acknowledge the use of CMBFAST of U. Seljak \& M. Zaldarriaga 
\cite{SelZal96}. We thank Wayne Hu for many useful discussions.

\end{document}